\documentstyle[12pt,fleqn]{article}
\begin{document}
\baselineskip=6.6mm

\begin{center}
{\large\bf GRAVITATIONAL WAVES FROM PHASE TRANSITIONS
                OF ACCRETING NEUTRON STARS}

\vspace{10.0mm}
K.S. Cheng$^1$ and Z.G. Dai$^2$

$^1${\em Department of Physics, University of Hong Kong, Hong Kong}

$^2${\em Department of Astronomy, Nanjing University, Nanjing 210093, 
China}
\end{center}

\vspace{10mm}

\begin{center}
ABSTRACT
\end{center}

We propose that when neutron stars in low-mass X-ray binaries accrete 
sufficient mass and become millisecond pulsars, the interiors of these stars 
may undergo phase transitions, which excite stellar radial oscillations.
We show that the radial oscillations will be mainly damped by 
gravitational-wave radiation instead of internal viscosity.
The gravitational waves can be detected by the advanced Laser Interferometer 
Gravitational-Wave Observatory at a rate of about three events per year.

\noindent
{\em Subject headings:} dense matter --- gravitation --- stars : neutron
\newpage

\begin{center}
1. INTRODUCTION
\end{center}

Gravitational wave astronomy may soon become an observational 
science, since three gravitational wave experiments, including 
the Laser Interferometer Gravitational-Wave Observatory 
(LIGO) (Abramovici et al. 1992), are under construct. In astrophysics, 
neutron stars are widely believed to be the most promising source of 
gravitational radiation (for detailed reviews see Thorne 1987, 1995), 
which from collapse of the 
cores of massive stars may provide a signature for the features of supernova
explosion (Burrows \& Hayes 1996), which from starquakes of pulsars 
may reveal the physics of the stellar interiors (Zimmermann \& Szedenits 
1979),  and which from mergers of binary neutron
stars may give information about the equation of state (EOS) of nuclear 
matter at high densities (Shibata, Nakamura \& Oohara 1992, 1993; 
Rasio \& Shapiro 1994; Zhuge, Centrella \& McMillan 1994; Davies et al. 
1994; Ruffert, Janka \& Sch\"afer 1996). 
In this Letter we propose a new possible 
origin of gravitational-wave bursts. We argue that when neutron stars 
in low-mass X-ray binaries accrete sufficient mass and become millisecond 
pulsars the interiors of these stars can undergo phase transitions, 
which excite stellar radial oscillations, producing strong 
gravitational wave bursts. 

\begin{center}
2. THE MODEL
\end{center}

\begin{center}
{\em 2.1. Evolution of Neutron Stars in Low-Mass X-Ray Binaries}
\end{center}

According to the standard scenario of evolution of
low-mass X-ray binaries (Bhattacharya \& van den Heuvel 1991), mass is 
transferred from the companion to the neutron star, which is spun up
to a millisecond period. The mass-transfer from the companion 
keeps driving the two stars apart while processes such as
orbital gravitational radiation or magnetic braking keeps driving 
the two stars closer. These processes keep the system in a steady 
mass-transfer state throughout the evolutionary timescale of the companion 
while the accretion rate of the neutron star is 
near the Eddington value. Thus, the neutron star
can accrete mass $\ge 0.5M_{\odot}$ in $\sim 10^8$ years and  
become a millisecond pulsar (van den Heuvel \& Bitzaraki 1995a, b). 
If we make an assumption that 
the masses of neutron stars before accretion are $1.4M_{\odot}$, which 
is supported by the current theories of Type II supernova explosion and
observations of masses of pulsars (e.g., the Hulse-Taylor binary system),
then the stars in an evolutionary timescale $\ge 10^8\,$yrs must
become rather massive ones ($\ge 1.8M_{\odot}$). Now we ask a question:
what could possibly occur in the interiors of these massive neutron stars?

To study this question, we first analyze possible EOSs for neutron stars.
So far there have been many approaches to determine an EOS for dense matter
through the many-body theory of interacting hadrons. Unfortunately, these 
approaches have given EOSs with different stiffnesses and in turn
very different structures of neutron stars. However, the EOSs should be 
constrained by the observations as follows. First, Link, Epstein \& van
Riper (1992) used a model-independent approach to analyze the postglitch 
recovery in four isolated pulsars (Crab, Vela, PSR\,0355+54 and 
PSR\,0525+21) which are likely to be isolated $1.4M_{\odot}$ neutron stars,
and concluded that soft EOSs at high densities are ruled out. 
More detailed analyses of the postglitch curves of the Crab and Vela 
pulsars also draw similar conclusions (Alpar et al. 1993, 1994). 
Second, if the EOSs in cores of neutron stars with mass $\sim 
1.4M_{\odot}$ were soft, the massive compact objects after the accretion 
phase of low-mass X-ray binaries could be black holes (Brown 1988). 
In fact, these objects have been identified as millisecond pulsars. 
This means that soft EOSs are unlikely to occur in neutron stars 
with mass $\sim 1.4M_{\odot}$. For these two reasons, we can assume 
that the EOS in neutron stars with mass $\sim 1.4M_{\odot}$ is 
moderately stiff to stiff.  

The above assumption is consistent with recent theoretical studies 
of the EOS for dense matter at high densities. First, because of the 
strong repulsion between nucleons and nucleon holes in the spin-isospin
interaction, pion condensation is unlikely in neutron stars (Brown et al. 
1988; Baym 1991).
Second, the possibility of kaon condensation in dense matter was suggested
by Kaplan and Nelson (1986), whose basic idea 
is that the energy of a negative kaon is lowered by 
interaction with nucleons. In neutron star matter in beta equilibrium
one expects negative kaons to be present if the energy to create one kaon 
in the matter is less than the electron chemical potential. 
However, Pandharipande, Pethick and Thorsson (1995) 
studied kaon-nucleon and nucleon-nucleon correlations
in kaon condensation in dense matter and found the kaon energy is much 
larger than that of the previous studies (see, e.g., Brown et al. 1992;
Brown et al. 1994; Thorsson, Prakash \& Lattimer 1994; Lee et al. 1995;
Thorsson \& Wirzba 1995). On the other hand, if we adopt 
electron chemical potentials of the calculations of the simple parametrized
models of Prakash, Ainsworth and Lattimer (1988) which represent a number
of more realistic models and compare these potentials with the kaon 
energies of the Hartree calculations (Pandharipande et al. 1995) in Figure 1, 
then we can see 
that the density for kaon condensation is $\sim$ (5-7)$\rho_{0}$ (where 
$\rho_{0}$ is the nuclear density), which significantly exceeds the central 
densities of neutron stars with mass $\sim 1.4M_{\odot}$ and with moderately 
stiff or stiff EOSs. Several authors (e.g., Ellis, Knorren \& Prakash 1995;
Schaffner \& Mishustin 1996; Dai \& Cheng 1996) have used the relativistic
mean-field approach to study kaon energies in neutron star matter, and
have also drawn similar conclusions. 
Third, it is thought that the density for deconfinement 
of nuclear matter to two-flavor quark matter is near (6-9)$\rho_{0}$ 
(Baym 1991). Therefore, the recent theoretical studies of dense matter
also indicate that pion (or kaon) condensation or quark matter is unlikely to
occur in neutron stars with mass $\sim 1.4M_{\odot}$.

We now turn to the study of what happens in neutron stars 
with moderately stiff to stiff EOSs when they accrete
mass $\ge 0.5M_{\odot}$, viz., when their masses increase from $1.4M_{\odot}$
to more than $1.8M_{\odot}$. It is clear that the central densities of
these massive stars are $\sim$\,(5-7)$\rho_{0}$. Once this condition is 
reached, several physical processes will take place in the stars. When 
$e^{-}\rightarrow K^{-}+\nu_{e}$, kaon condensation occurs spontaneously 
in the stellar interior. The appearance of this new phase destroys 
the hydrostatic equilibrium of the star due to the softening of the EOS
of the core matter. This implies a structural transition
of the whole star into a new, stable configuration with a condensed core
of the new phase. The structural transition --- neutron star 
corequake --- occurs on the dynamic timescale of milliseconds, in which
the interior temperature may increase to $\sim 10$\,MeV 
both due to the rapid compression of dense matter (Haensel, Denissov 
\& Popov 1990) and due to 
the energy released by reaction $e^{-}\rightarrow K^{-}+\nu_{e}$. 
Alternatively, if the central nuclear matter 
is deconfined into two-flavor quark matter, the quark matter will convert
to three-flavor quark matter because strange matter is thought to be more
stable than nuclear matter. Thus a strange matter seed is formed in the 
interior, and subsequently the strange matter will begin to swallow the
neutron matter in the surroundings. This process should proceed in a 
timescale $\sim$ tens of milliseconds due to a detonation mode, and 
the interior temperature increases to $\ge 10\,$MeV because the chemical
energy of the two-flavor quark matter is dissipated into thermal 
energy (Dai, Peng \& Lu 1995). The conversion of neutron stars 
to strange stars has been suggested as a possible origin of 
cosmological $\gamma$-ray bursts (Cheng \& Dai 1996). Furthermore, the
phase transitions of massive neutron stars to either stars with kaon 
condensation cores or strange stars can stimulate stellar radial 
oscillations. In next subsection we will show that these newly born 
compact objects may be a strong source of gravitational-wave radiation 
if their rotation periods are of the order of milliseconds.

\begin{center}
{\em 2.2. Damping of Radial Oscillations}
\end{center}

Radial oscillations are damped not only due to dissipation of the vibration
energy of stellar matter into heat but also due to conversion of this energy 
into gravitational-wave radiation. We first focus on the case in which 
a massive neutron star undergoes a phase transition to a star containing 
a kaon condensation core. After $e^{-}\rightarrow K^{-}+\nu_{e}$, electron
neutrinos are trapped in the interior and form an ideal Fermi-Dirac
gas with $\mu_{\nu}\gg kT$ (where $T$ is the interior temperature) because 
the neutrino mean free path is much less than the stellar radius for  
$KT \sim 10\,$MeV. Subsequent evolution of the newborn star, which
is analogous to that of a protoneutron star formed from supernova explosion
(Burrows \& Lattimer 1986), can be divided into three stages. 
We neglect gravitational radiation. The first stage is deleptonization, 
whose timescale ($\tau_{d1}$) is of the order of 1\,s (Sawyer \& Soni 1979), 
in which the interior temperature may increase significantly. In this stage
the radial oscillations are damped through the following reaction 
\begin{equation}
\tilde p + e^{-} \leftrightarrow \tilde n + \nu_{e}\,, 
\end{equation}
where $\tilde p$ and $\tilde n$ represent quasiprotons and quasineutrons 
respectively. Using the reaction matrix element of Brown et al. (1988),
we have derived the net rate per baryon for this reaction at nonequilibrium,
\begin{equation}
\Gamma_{1}=\frac{1}{24\pi^5\hbar^{10}c^5}G_{F}^2\cos^2\theta_{C}
\cos^2(\theta/2)(1+3g_A^2)m_n^*m_p^*\mu_e\mu_{\nu}^2\Delta\!\mu 
[\Delta\!\mu^2+(2\pi kT)^2]\,,
\end{equation}
where $G_F$ is the weak coupling constant, $\theta_C$ is the Cabibbo angle,
$\theta$ is the chiral angle for kaon condensation ($\theta \le 60^{\rm o}$) 
(Brown et al. 1994),   
$g_A$ is the Gamow-Teller coupling constant, $m_n^*$ and $m_p^*$ are the
effective nucleon masses, and $\Delta\!\mu=\mu_p+\mu_e-\mu_n-\mu_\nu$ with 
$\mu_i$ being the chemical potential of particle $i$. 
In deriving equation (2), we have assumed $kT/c\ll |\!{\bf P}_\nu
\!|\ll |{\bf P}_n\!|$ with ${\bf P}_\nu$ 
and ${\bf P}_n$ being the neutrino and neutron momenta respectively.
As the steps shown by Dai \& Lu (1996), using this reaction rate, 
we have further derived the bulk viscosity as
\begin{equation}
\eta_{1} \simeq 1.8\times 10^{25} \cos^{-2}(\theta/2)Y_{e}^{-1/3}Y_{\nu}
^{2/3}Y_{n}^{4/3}\left(\frac{\rho}{\rho_{0}}\right)^{5/3}\left(\frac{kT}
{1{\rm MeV}}\right)^{-2}\,\,{\rm g}\,{\rm cm}^{-1}\,{\rm s}^{-1}\,,
\end{equation}
where $Y_{e}$, $Y_{\nu}$ and $Y_{n}$ are the particle concentrations,
and $\rho$ is the stellar density. Therefore, the damping timescale 
(Sawyer 1980) is given by
\begin{equation}
\tau_{v1}  =  \frac{1}{30}\rho R^2\eta_1^{-1} 
\end{equation}
where $R$ is the stellar radius. Inserting equation (3) into this equation,  
we obtain
\begin{equation}
\tau_{v1} \simeq 0.52\cos^2 (\theta/2)Y_{e}^{1/3}Y_{\nu}^{-2/3}
Y_{n}^{-4/3}R_6^2\left(\frac{\rho}{\rho_{0}}\right)^{-2/3}
\left(\frac{kT}{1{\rm MeV}}\right)^2\,\,{\rm s}\,,
\end{equation}
where $R_6$ is in units of $10^6\,$cm.
For the typical particle concentrations and $\rho\sim 6\rho_{0}$,
$R\sim 10^6\,{\rm cm}$ and $kT\sim 10\,$MeV, $\tau_{v1}\sim
12Y_e^{1/3}Y_\nu^{-2/3}Y_n^{-4/3}\,{\rm s}>12\,{\rm s}\gg \tau_{d1}$. 

In the second stage, the stellar interior has 
practically no trapped lepton-number excess, as compared with 
catalyzed matter. The diffusion of $\nu_e\bar {\nu}_e$ is 
then driven by the temperature gradient. Locally, 
the equilibrium distribution function of $\nu_e\bar {\nu}_e$ can be 
approximated by the Fermi-Dirac one with a zero chemical potential.
The neutrino diffusion timescale ($\tau_{d2}$) is of the order of $40\,$s 
(Sawyer \& Soni 1979). Using the analogy of deriving the bulk viscosity 
of (Haensel \& Zdunik 1992), we obtain the damping timescale 
\begin{equation}
\tau_{v2}\simeq 1.8\times 10^2\cos^2 (\theta/2)Y_e^{-1/3}R_6^2
\left(\frac{\rho}{\rho_{0}}\right)^{2/3}\,\,{\rm s}\,,
\end{equation}
Clearly, $\tau_{v2}\gg \tau_{d2}$. 

Third, after the neutrino diffusion, 
the temperature decreases to $\sim 1\,$MeV, and neutrinos can escape
freely from the star. The damping timescale becomes 
\begin{equation}
\tau_{v3}\simeq 2.6\times 10^{-5}\cos^{-2} (\theta/2)Y_{e}^{-1/3}
R_6^2\left(\frac{\rho}{\rho_{0}}\right)^{2/3}
\left(\frac{kT}{1{\rm MeV}}\right)^{-4}
\left(\frac{\omega}{10^4{\rm s}^{-1}}\right)^2\,\,{\rm s}\,,
\end{equation}
where $\omega$ is the oscillation frequency.
Therefore, we can conclude that the radial oscillations are not
damped very efficiently by the bulk viscosity until the stage at which 
neutrinos escape freely. 
                           
We now consider gravitational radiation from rapidly spinning and 
oscillating neutron stars. The timescale for this process (Chau 1967) is
\begin{equation}
\tau_{g}\simeq 0.41 M_{1.8}^{-1}
R_6^{-2}\left(\frac{P}{2{\rm ms}}\right)^4\,\,{\rm s}\,,
\end{equation}
where $M_{1.8}$ is the stellar mass in units of $1.8M_\odot$, and $P$ 
is the stellar rotation period. 
Here we have assumed that nucleons in the stellar interior are 
nonrelativistic and degenerate, and thus the adiabatic index 
is equal to 5/3. When $\tau_g\le \tau_{d1}+\tau_{d2}$, viz., 
\begin{equation}
P\le 6.2\,\bar{\tau}_d^{1/4} M_{1. 8}^{1/4} R_6^{1/2} \,\,{\rm ms}\,, 
\end{equation}
where $\bar{\tau}_d=(\tau_{d1}+\tau_{d2})/40{\rm s}$,
the gravitational radiation can damp the radial oscillations very 
efficiently. The frequency of the gravitational waves 
is euqal to $\omega=2\pi/\tau$, where $\tau$ is close 
to $5\times 10^{-4}\,$s for a typical neutron-star model 
(Glass \& Lindblom 1983) (corrected
for the gravitational redshift) with mass $M\sim 1.8M_{\odot}$ and
radius $R\sim 10^6\,$cm. The strength of the waves can be estimated
using the quadrupole approximation to the Einstein field equations (Thorne 
1987). This approximation shows that the gravitational strain is given by
\begin{equation}
h \simeq \frac{G}{c^4}\frac{\ddot Q}{r}
\sim 1.5\times 10^{-23}R_6^5\left(\frac{\rho}{6\rho_{0}}
\right)\left(\frac{\alpha}{0.1}\right)
\left(\frac{P}{2{\rm ms}}\right)^{-2}\left(\frac{r}{100{\rm Mpc}}
\right)^{-1}\,,
\end{equation}
where $Q$ is the source's quadrupole moment (Chau 1967), $r$ is the distance 
of the source from Earth, and $\alpha$ is the relative oscillation amplitude
($=\delta R/R$). Particularly, in the case of phase transitions of 
neutron stars with stiff EOSs to stars with kaon condensation cores, 
$\alpha$ is expected to be about 0.1. 
Furthermore, the characteristic gravitational strain is
\begin{equation}
h_{c} \simeq h\sqrt{n}\sim 4.3\times 10^{-22} M_{1.8}^{-1/2}R_6^4
\left(\frac{\rho}{6\rho_{0}}\right)
\left(\frac{\alpha}{0.1}\right)\left(\frac{\tau}{0.5{\rm ms}}\right)^{-1/2}
\left(\frac{r}{100{\rm Mpc}}\right)^{-1}\,,
\end{equation}
where $n$ is the number of cycles of gravitational waves 
in the duration $\sim \tau_{g}$. The observed number of 
low-mass X-ray binaries in our Galaxy is $\sim 10^2$ (van Paradijs 1995), 
and thus the rate for phase transitions of their neutron stars
is $\sim 10^{-6}\,{\rm yr}^{-1}$ because the typical accretion timescale
is $\sim 10^8$ years. An alternative estimation based 
on the number of millisecond pulsars and their life time also 
gives a similar value (Cheng \& Dai 1996). Therefore, the rate 
detected by the advanced LIGO detector (Abramovici et al. 1992) 
is estimated to be about three events per year.

Alternatively, neutron stars in low-mass X-ray binaries may accrete 
sufficient mass to convert into strange stars, 
as suggested by Cheng \& Dai (1996). Next we would discuss this case.
First, the timescale for damping radial oscillations due to 
bulk viscosity is $\ge 10$\,s for a high temperature $\ge 10\,$MeV
(Madsen 1992; Dai \& Lu 1996), so 
the gravitational radiation damping mechanism is also more efficient
in this case. Second, for the conversion of neutron stars with stiff EOSs 
to strange stars, the relative oscillation amplitude ($\alpha$) 
may not be less than 0.1. Third, the density
for deconfinement may be larger than that for kaon condensation, and thus
the oscillation frequency of strange stars is larger. If 
gravitational waves discussed in this work are observed by the advanced 
LIGO in the future, then observations, in principle, can distinguish between
these two kinds of phase transitions. In addition, gravitational radiation 
for the case in which neutron stars convert to strange stars is likely 
to occur together with cosmological $\gamma$-ray bursts, because  
the fireballs formed during the conversion have very low baryon 
contamination (Cheng \& Dai 1996).

\begin{center}
3. CONCLUSIONS
\end{center}

We in this Letter have suggested that during the evolution of neutron stars
in low-mass X-ray binaries the stars may undergo phase transitions to stars 
containing kaon condensation cores or strange stars when these neutron 
stars accrete sufficient mass from their companions. The phase transitions
can excite stellar radial oscillations, which produce strong 
gravitational wave bursts if the stellar rotation periods are of the order 
of milliseconds. The study of such gravitational radiation 
may provide information for the high-density EOS and the physics 
of phase transitions. In addition, the rate detected by 
the advanced LIGO is estimated to be about three events per year.

Gravitational waves from colliding 
neutron stars must have a continuous spectrum, but waves from phase 
transitions of neutron stars appear to have a delta-function like spectrum. 
This signature is expected to be confirmed by very near future 
observations .

\vspace{5mm}
We thank W.M. Suen for bringing this problem to
our attention, W.Y. Chau for discussions on gravitational radiation,
and Samuel Wong and M.C. Chu for discussions on equations of state
at high densities. K.S.C. thanks a RGC grant of Hong Kong for support;
Z.G.D. thanks the National Natural Science Foundation of China for support.

\newpage
\baselineskip=4mm

\begin{center}
REFERENCES
\end{center}

\begin{description}
\item  Abramovici, A., et al. 1992, Sci, 256, 325
\item  Alpar, M.A., Chau, H.F., Cheng, K.S., \& Pines, D. 1993, ApJ, 
           409, 345 
\item  --------------. 1994, ApJ, 427, L29
\item  Baym, G. 1991, in Neutron Stars: Theory and Observation, eds.
           J. Ventura and D. Pines, (Kluwer, Dordrecht), 21
\item  Bhattacharya, D., \& van den Heuvel, E.P.J. 1991, Phys. Rep., 203, 1
\item  Brown, G.E. 1988, Nat, 336, 519
\item  Brown, G.E., Kubodera, K., Page, D., \& Pizzochero, P. 1988, 
           Phys. Rev. D, 37, 2042
\item  Brown, G.E., Kubodera, K., Rho, M., \& Thorsson, V. 1992, 
           Phys. Lett. B, 291, 355  
\item  Brown, G.E., Lee, C.-H., Rho, M., \& Thorsson, V. 1994, 
           Nucl. Phys. A, 567, 937
\item  Burrows, A., \& Hayes, J. 1996, Phys. Rev. Lett., 76, 352
\item  Burrows, A., \& Lattimer, J.M. 1986, ApJ, 307, 178
\item  Chau, W.Y. 1967, ApJ, 147, 667
\item  Cheng, K.S., \& Dai, Z.G. 1996, Phys. Rev. Lett., 77, 1210 
\item  Dai, Z.G., \& Cheng, K.S. 1996, Phys. Lett. B, submitted
\item  Dai, Z.G., \& Lu, T. 1996, Z. Phys. A, 355, 415
\item  Dai, Z.G., Peng, Q.H., \& Lu, T. 1995, ApJ, 440, 815 
\item  Davies, M.B., Benz, W., Piran, T., \& Thielemann, F.K. 1994, 
          ApJ, 431, 742  
\item  Ellis, P.J., Knorren, R., \& Prakash, M. 1995, Phys. Lett. B, 349, 11
\item  Glass, E.N., \& Lindblom, L. 1983, ApJS, 53, 93
\item  Haensel, P., Denissov, A., \& Popov, S. 1990, A\&A, 240, 78
\item  Haensel, P., \& Shaeffer, R. 1992, Phys. Rev. D, 45, 4708
\item  Kaplan, D.B., \& Nelson, A.E. 1986, Phys. Lett. B, 175, 57
\item  Lee, C.-H., Brown, G.E., D.-P. Min, D.-P., \& Rho, M. 1995, 
           Nucl. Phys. A, 585, 401 
\item  Link, B., Epstein, R.I., \& Van Riper, K.A. 1992, Nat, 359, 616 
\item  Madsen, J. 1992, Phys. Rev. D, 46, 3290 
\item  Pandharipande, V.R., Pethick, C.J., \& Thorsson, V. 1995,
           Phys. Rev. Lett., 75, 4567
\item  Prakash, M., Ainsworth, T.L., \& Lattimer, J.M. 1988, 
           Phys. Rev. Lett., 61, 2518 
\item  Rasio, F.A., \& Shapiro, S.L. 1994, ApJ, 432, 242 
\item  Ruffert, M., Janka, H.-T., \& Sch\"afer, G. 1996, A\&A, 311, 532 
\item  Sawyer, R.F. 1980, ApJ, 237, 187
\item  Sawyer, R.F., \& Soni, A. 1979, ApJ, 230, 859
\item  Shaffner, J., \& Mishustin, I.N. 1996, Phys. Rev. C, 53, 1416
\item  Shibata, M., Nakamura, T., \& Oohara, K. 1992, Prog. Theor. Phys., 
           88, 1079 
\item  --------------. 1993, Prog. Theor. Phys., 
           89, 809 
\item  Thorne, K.S. 1987, in Three Hundred Years of Gravitation, eds.
          S.W. Hawking \& W. Israel, (Cambridge Univ. Press), 330.
\item  --------------. 1995, in Proceedings of the Snowmass 95 Summer Study 
          on Particle and Nuclear Astrophysics and Cosmology, eds. E.W. Kolb 
          \& R. Peccei, (World Scientific, Singapore), 1
\item  Thorsson, V., Prakash, M., \& Lattimer, J.M., Nucl. Phys. 
           A, 572, 693 
\item  Thorsson, V., \& Wirzba, A. 1995, Nucl. Phys. A, 589, 633
\item  van den Heuvel, E.P.J., \& Bitzaraki, O. 1995a, A\&A, 297, L41
\item  --------------. 1995b, in The Lives of the Neutron Stars, eds. 
           M.A. Alpar, \"U. Kilziloglu \& J. van Paradijs (Kluwer, Dordrecht)
\item  van Paradijs, J. 1995, in X-Ray Binaries, eds. W.H.G. Lewin, J. van 
           Paradijs and E.P.J. van den Heuvel, (Cambridge: Cambridge
           Univ. Press), 536
\item  Zhuge, X., Centrella, J.M., \& McMillan, S.L.M. 1994, Phys. Rev.
          D, 50, 6247 
\item  Zimmermann, M., \& Szedenits, E. 1979, Phys. Rev. D, 20, 351 
\end{description}

\newpage

\baselineskip=8mm

\begin{center}
FIGURE CAPTION
\end{center}

\noindent
FIG. 1. Energy of a single negative kaon 
in neutron star matter and electron chemical potential 
as functions of density. The three solid lines are kaon energies from 
the Hartree calculations of Pandharipande et al. (1995) for square wells 
of radii $R=1$ 
and 0.7\,fm, and for a Yukawa potential. The dashed lines labeled by PAL1,
PAL2 and PAL3 represent electron chemical potentials calculated 
from the parametrized models of Prakash et al. (1988) corresponding to three 
different forms of $F(u)$ which parameterizes the potential 
contribution to the symmetry energy.

\end{document}